\documentclass[useAMS,usenatbib]{mn2e}

\usepackage{BibDef}
\usepackage{hyperref}
\usepackage{graphicx}
\usepackage{amssymb}

\title[Simulations of CRGs: interaction-geometry effects]{
Adaptive mesh refinement simulations of collisional ring galaxies: effects of the interaction geometry
}

\author[D. Fiacconi et al.]{Davide Fiacconi$^{1}$\thanks{E-mail: davide.fiacconi@gmail.com}, Michela Mapelli$^{2}$, Emanuele Ripamonti$^{1}$ and Monica Colpi$^{1}$\\
$^{1}$Universit\`a di Milano-Bicocca, Dipartimento di Fisica ``G. Occhialini'', Piazza della Scienza 3, I-20126, Milano, Italy\\
$^{2}$INAF-Osservatorio Astronomico di Padova, Vicolo dell'Osservatorio 5, I-35122, Padova, Italy
}

\begin{document}

\date{\today}

\pagerange{\pageref{firstpage}--\pageref{lastpage}} \pubyear{2012}

\maketitle

\label{firstpage}


\begin{abstract}
Collisional ring galaxies are the outcome of nearly axisymmetric high-speed encounters between a disc and an intruder galaxy.
We investigate the properties of collisional ring galaxies as a function of the impact parameter, the initial relative velocity and the inclination angle.
We employ new adaptive mesh refinement simulations to trace the evolution with time of both stars and gas, taking into account star formation and supernova feedback.
Axisymmetric encounters produce circular primary rings followed by smaller secondary rings, while off-centre interactions produce asymmetric rings with displaced nuclei.
We propose an analytical treatment of the disc warping induced by an inclination angle greater then zero.
The star formation history of our models is mainly influenced by the impact parameter: axisymmetric collisions induce impulsive short-lived starburst episodes, whereas off-centre encounters produce long-lived star formation.
We compute synthetic colour maps of our models and we find that rings have a $B-V$ colour typically $\sim 0.2 \; {\rm mag}$ bluer than the inner and outer disc, in agreement with observations.
\end{abstract}



\section{Introduction}

Collisional ring galaxies (CRGs) consist of one or more sharply defined star-forming rings, generally (but not always) with a nucleus inside \citep*{madore+09}.
Theoretical models, both analytical (e.g. \citealt{struck-marcell+90}; \citealt{lotan+90}; \citealt{gerberlamb94}; \citealt{appleton+96}) and numerical (e.g. \citealt{lynds+76}; \citealt{theys+77}; \citealt{appleton+87}; \citealt{appleton+90}; \citealt{hernquist+93}; \citealt{struck-marcell+93};  \citeauthor*{gerber+94} \citeyear{gerber+94}, \citeyear{gerber+96}; \citealt{horellou+01}; \citealt*{donghia+08}; \citealt{ghosh+08}; \citeauthor{mapelli+08} \citeyear{mapelli+08}, \citeyear{mapelli+08b}; \citealt{mapelli+12}; \citealt{smith+12}), suggest that CRGs are the result of nearly axisymmetric high-speed encounters between a disk galaxy (the `target' or `primary' galaxy) and an intruder galaxy.
After the interaction, radially expanding density waves form because of the crowding of star orbits in the target disk.
This causes gas compression and the triggering of a starburst episode along the expanding ring (e.g. \citealt{higdon+95}; \citealt{higdon+97}; \citealt{mayya+05}; \citealt*{bizyaev+07}; \citealt{rappaport+10}; \citealt{fogarty+11}).

CRGs are unique laboratories for the study of galaxy collisions because of their simple interaction geometry.
For example, the kinematics of the ring provides information about the dynamics of the interaction (e.g. \citealt{fosbury+77}; \citealt*{few+82}; \citealt{charmandaris+94}; \citealt{bizyaev+07}; \citealt{fogarty+11}).
This information is used to constrain $N$-body/hydrodynamical models reproducing the formation process of well studied CRGs such as the Cartwheel galaxy (e.g. \citealt{hernquist+93}; \citealt{struck-marcell+93}; \citealt{horellou+01}; \citealt{mapelli+08}) and Arp 147 (e.g. \citealt*{gerber+92}; \citealt{mapelli+12}).

Numerical simulations are helpful tools to explore the parameter space of the collision and to disentangle the effect of each parameter on the evolution of CRGs.
\citet{gerber+96} study the properties of CRGs formed in axisymmetric encounters with different intruder-to-target mass ratios.
They show that, as the intruder mass increases, the ring becomes sharper, expands more rapidly and the secondary ring is less developed.
\citet{mapelli+12} explore the connection between gas fraction and star formation (SF) in simulations of empty CRGs (i.e. CRGs without a central nucleus, such as Arp 147), and find that the gas fraction strongly influences the peak of SF rate (SFR) after the starburst.
\citet{smith+12} perform a large parameter study with the aim of reproducing the dynamical properties of the Auriga's Wheel \citep{conn+11}, but they do not take into account the effects of the interaction parameters on the SF.

In this paper, we investigate how the geometry of the impact affects  the morphology and SF history of CRGs. 
We employ a suite of adaptive mesh refinement (AMR) numerical simulations to study systematically the properties of CRGs as a function of the interaction parameters that define the geometry of the encounter. 
This is the first time that the AMR technique is applied to a wide grid of simulations studying the formation of CRGs.
Our choice is particularly important, as the AMR technique ensures a better treatment of hydrodynamical instabilities and shocks with respect to particle-based schemes (e.g. \citealt{agertz+07}; \citealt{price+08}). 
The paper is organized as follows: in Section \ref{section2} we describe the main characteristics of our suite of simulations and briefly discuss the parameters of interest.
In Section \ref{section3} we present the results, and focus on the morphology and SF history of our models.
We also extract synthetic photometric information about the colour of the ring and we compare it with observations. 
In Section \ref{section4} we summarize our main findings.


\section{Numerical simulations}\label{section2}

\subsection{Numerical code} \label{subsec21}

Our simulations of galaxy interactions were performed with the public AMR code RAMSES \citep{teyssier+02}.
RAMSES solves the fluid equations on an adaptive Cartesian grid by means of a second-order Godunov method coupled with a Riemann solver.
This method ensures a better treatment of hydrodynamical instabilities and shocks with respect to particle-based schemes (e.g. \citealt{agertz+07}; \citealt{price+08}).
Gravitational interactions are calculated using Particle Mesh (PM) techniques for $N$-body collisionless components (stars and dark matter particles), solving the Poisson equation on the AMR grid with the multigrid strategy of \citet{guillet+11}.

All the simulations presented in this paper adopt a cubic box of 960 kpc per edge, with a static coarse grid composed of $64^3$ cells.
We used 7 additional levels of refinement to reach the maximum resolution of $\Delta x \simeq 117$ pc.
Grid refinement is triggered following three criteria, one for $N$-body particles, the other two for the gaseous component.
The particle criterion requires that at least 24 particles reside in a cell before the refinement, so as to maintain a roughly constant number of particles per cell.
The first criterion for gas is based on a `quasi-lagrangian' approach according to which refining is triggered if the gas mass inside a cell exceeds a threshold of $\sim 1.5 \times 10^{5} \; {\rm M_{\odot}}$.
The second gas criterion is based on the Jeans length (\citealt{jeans19}): a cell at level $\ell$ is split if the local Jeans length does not exceed 4 times the cell dimension $\Delta x^{(\ell)}$.
In the limit of our maximum resolution, this choice allows us to properly follow the evolution of features such as dynamically unstable gaseous clumps in the expanding ring.
Gas dynamics is followed assuming a polytropic equation of state (EOS) with adiabatic index $\gamma = 5/3$, corresponding to a perfect mono-atomic gas.
The interacting galaxies are embedded in a hot and thin medium with numerical density $n_{\rm bg} \simeq 10^{-6} \; {\rm cm^{-3}}$, and temperature $T_{\rm bg} = 5 \times 10^{5} \; {\rm K}$.

The RAMSES code includes physically-motivated sub-grid models for gas cooling, SF and supernova (SN) feedback.
The code adopts the cooling function described in \citet{sutherland+93} for temperatures down to $\sim 10^4 \; {\rm K}$.
We include a sub-grid gaseous EOS $T = T_{0} (\rho/\rho_0)^{\gamma_0 - 1}$ to model the behavior of the inter-stellar medium (ISM), where $T_0 = 1000 \; {\rm K}$, $\gamma_0 = 5/3$ and $\rho_0$ is the SF density threshold (see \citealt*{agertz+11}).
SF is implemented in RAMSES following the prescriptions by \citet{rasera+06}.
The conversion of gas density $\rho_{\rm gas}$ into star density $\rho_{\star}$ is based on a Schmidt-law \citep{schmidt+59} of the form:
\begin{equation}
\frac{{\rm d} \rho_{\star}}{{\rm d}t} = \epsilon_{\rm SF} \frac{\rho_{\rm gas}}{t_{\rm dyn}} \qquad {\rm if} \qquad \rho_{\rm gas} > \rho_{0},
\end{equation}
where $\epsilon_{\rm SF}$ is the SF efficiency, $t_{\rm dyn}$ is the local dynamical time and $\rho_0$ is a density threshold.
We chose $\epsilon_{\rm SF} = 0.1$ and $\rho_0$ corresponding to a numerical density  $n_0=5 \; {\rm cm^{-3}}$.
Such threshold density was selected to resolve the Jeans length (\citealt{jeans19}) for a temperature $T\sim{}10^3$ K (i.e. consistent with the minimum gas temperature in our simulations) with our best spatial resolution (see \citealt{mapelli+12} for details).
Thus, the adopted threshold density is the highest that can be safely used with our spatial resolution and matches the one adopted in recent cosmological simulations that reproduce realistic Milky Way-like (MW-like) galaxies \citep{guedes+11}, and in simulations of the formation of off-centred CRGs \citep{mapelli+12}. 
Most of cosmological simulations adopt a much lower density threshold ($n_0=0.1\; {\rm cm}^{-3}$, e.g. \citealt{agertz+11} and references therein), that might unphysically enhance the SF (see e.g. the discussion in \citealt{guedes+11}).
\citet{governato+10} adopt $n_0=100 \; {\rm cm}^{-3}$ (a factor of 20 higher than our threshold) in their cosmological simulations, but their spatial resolution is better than ours and their cooling function goes down to $\sim{}300 \; {\rm K}$. 

Our adopted value of the SF efficiency ($\epsilon_{\rm SF} = 0.1$) is a factor of $2-10$ higher than adopted in recent simulations (e.g. \citealt{agertz+11}), which follow the observational hints by \citet{krumholztan07}.
On the other hand, we stress that (i) the $\epsilon_{\rm SF}$ parameter in our simulations is simply phenomenological and does not have a direct relation to the true SF efficiency within molecular clouds (which we cannot simulate realistically, see the discussion below), and that (ii) \citet{guedes+11} show that the simulated SF is much more affected by $\rho_0$ than by $\epsilon_{\rm SF}$.

Although the SF recipe used by \citet{mapelli+12} is more elaborate than ours, we find similar results about the global SF history for qualitatively similar runs (see the discussion below).
Star particles are then spawned following a stochastic approach \citep{katz+92}, with a mass that is an integer multiple of $m_{\star}^{\rm (min)} = \rho_{0} \Delta x^3$.
We stress that our best grid resolution (117 pc) and our adopted cooling function \citep{sutherland+93} are not suitable to follow the formation of low-temperature ($<10^3$ K) high-density ($>10^2 \; {\rm cm}^{-3}$) gas clouds, whose typical scale is expected to be $\lesssim{}50$ pc and that were studied in previous simulations of multiphase interstellar medium (e.g. \citealt{gerritsenicke97}; \citealt{wada+02}; \citealt{bottema03}; \citealt{taskerbryan06}; \citealt{WadaNorman07}; \citealt{agertz+09}).
Therefore, our simulations do not model SF in cold clumps, which is physically the main channel for SF.
Accounting for this would be computationally prohibitive to do for a wide set of galaxy interaction runs.
SN feedback is implemented as thermal feedback due to Type II SN explosions characterized by a specific energy $\mathcal{E} \simeq 5 \times 10^{16} \; {\rm erg \, g^{-1}}$ \citep{agertz+11}.
A total energy amount $E = \epsilon_{\rm SN} \, m_{\star} \, \mathcal{E}$ is deposited in the 27 cells around the star particle of mass $m_{\star}$ during $\sim 10 \; {\rm Myr}$ after the star particle creation.
We assume a SN efficiency $\epsilon_{\rm SN} = 0.1$.


\subsection{Galaxy models}

We model the target galaxy as a MW-like disc galaxy composed of three $N$-body components and a gridded one: a dark matter (DM) halo, a stellar disc, a stellar bulge and a gaseous disc, respectively.
For the DM halo we choose the density profile by \citeauthor*{navarro+96} (\citeyear{navarro+96}, NFW):
\begin{equation}
\rho_{\rm h} (r) = \rho_{\rm crit} \, \delta_{\rm c} \, \frac{1}{(r/r_{\rm s}) (1 + r/r_{\rm s})^2},
\end{equation}
where $\rho_{\rm crit}$ is the critical density of the Universe\footnote{$\rho_{\rm crit} = 3H_0^2/(8 \pi G)$, where we assume a Hubble constant $H_0 = 71 \; {\rm km \, s^{-1} \, Mpc^{-1}}$ \citep{larson+11}.}, $r_{\rm s}$ is the halo scale radius and $\delta_c$ is defined as:
\begin{equation}
\delta_{\rm c} = \frac{200}{3} \, \frac{c^3}{\ln(1+c) - c/(1+c)},
\end{equation}
where $c$ is the NFW concentration, $c \equiv R_{200}/r_{\rm s}$, where $R_{200}$ is the virial radius of the DM halo so that the virial mass of the NFW halo is $M_{200}=   200 \, \rho_{\rm crit} \, (4/3) \, \pi \, R_{200}^3$.
We assume an axisymmetric exponential profile to model both the gaseous and stellar discs:
\begin{equation}
\rho_{\rm d} (R,z) = \frac{M_{\rm d}}{4 \pi R_{\rm d}^2 z_{\rm d}} \, e^{-R/R_{\rm d}} \, {\rm sech}^2 (z/z_{\rm d}),
\end{equation}
where $M_{\rm d}$ is the total disc mass, and $R_{\rm d}$ and $z_{\rm d} = 0.1 R_{\rm d}$ are the radial and vertical scale length of both the gaseous and stellar discs, respectively.
The gaseous disc has an initial constant temperature $T_0 = 15000 \; {\rm K}$ and we assume that it is composed of a mixture of ionized hydrogen and helium with a mean molecular weight $\mu \simeq 0.59$.
The density profile of the stellar bulge follows the spherical \citet{hernquist+90} model:
\begin{equation}
\rho_{\rm b} (r) = \frac{M_{\rm b}}{2 \pi a_{\rm s}^3} \, \frac{1}{(r/a_{\rm s}) (1+r/a_{\rm s})^3},
\end{equation}
where $M_{\rm b}$ is the total mass and $a_{\rm s}$ the radial scale length of the bulge.

The intruder galaxy is a dwarf, spherical early-type galaxy composed of a NFW DM halo and a central stellar bulge devoid of any gaseous component.
We chose not to simulate a gas-rich intruder because, in this paper, we are interested in studying the evolution of the gas of the target separately. 

The parameters for the initial conditions (ICs) of the two galaxies are shown in Table \ref{table1}.
\begin{table}
\caption{Adopted values of the model parameters for the target and intruder galaxy.
These parameters do not change among runs.}
\begin{center}
\label{table1}
\begin{tabular}{l c c}
\hline
Parameter & Target & Intruder \\
\hline
$M_{\rm h} \; [{\rm M_{\odot}}]$ & $1.16 \times 10^{12}$ & $5.9 \times 10^{11}$ \\
$r_{\rm s} \; [{\rm kpc}]$ & 20.0 & 12.0 \\
$c$ & 11.0 & 14.5 \\
$M_{\rm b} \; [{\rm M_{\odot}}]$ & $1.5 \times 10^{10}$ & $3.8 \times 10^{10}$ \\
$a_{\rm s} \; [{\rm kpc}]$ & 0.7 & 0.6 \\
$M_{\rm d} \; [{\rm M_{\odot}}]$ & $6.0 \times 10^{10}$ & - \\
$R_{\rm d} \; [{\rm kpc}]$ & 3.5 & - \\
$f_{\rm gas}^{\dag}$ & 0.075 & - \\
\hline
\end{tabular}

\medskip

$^{\dag}$ Fraction of gas with respect to the total disc mass, $f_{\rm gas} = M_{\rm gas} / (M_{\rm gas}+M_{\star})$.
\end{center}
\end{table}
ICs are generated by using the code described in \citet*{widrow+08}.
The code generates self-consistent disc-bulge-halo galaxy models derived from explicit distribution functions for each component (see also \citealt{kuijken+95}; \citealt{widrow+05}).
We use $\sim 1.5 \times 10^6$ particles for the DM halo of the target galaxy, while the target disc and bulge are composed of $\sim 3.7 \times 10^{5}$ and $\sim 1 \times 10^{5}$ particles, respectively.
The intruder DM halo is composed of $\sim 7.8 \times 10^{5}$ particles, while its stellar bulge has $\sim 2.5 \times 10^{5}$ particles.

The intruder-to-target mass ratio is $\sim 1/2$ for both the dark and baryonic components. 
Such a high value induces a strong perturbation on the target disc \citep{gerber+96}.
This choice is consistent with observations, indicating that the candidate intruder-to-target mass ratio is generally $\gtrsim 0.2$ (e.g. \citealt{appleton+96}).
We adopt a fraction of gas with respect to the total disc mass $f_{\rm gas} = M_{\rm gas} / (M_{\rm gas}+M_{\star}) = 0.075$, consistent with observations of MW-sized late-type galaxies in the local Universe (\citealt{zhang+09}, \citealt{evoli+11}).
We do not explore the effects of varying $f_{\rm gas}$ (see \citealt{mapelli+12} for this issue).


\subsection{Interaction parameters}

\begin{figure}
\includegraphics[width=80mm]{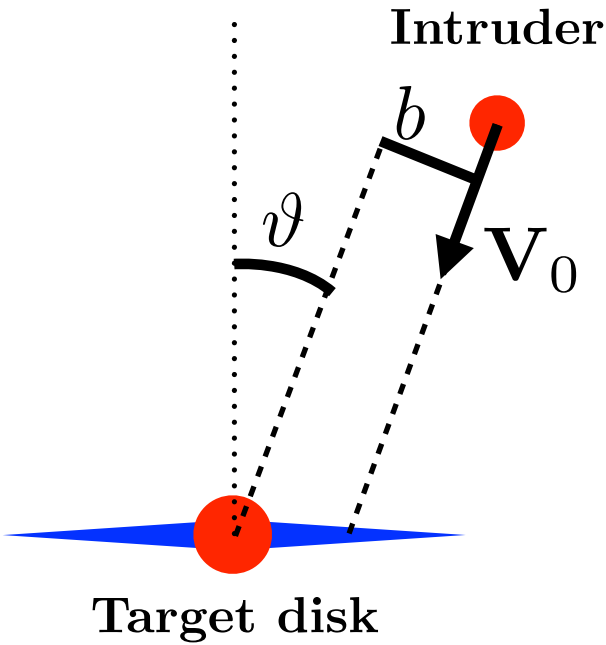}
\caption{Schematic representation of the ICs, showing $\mathbf{V}_0$, $b$ and  $\vartheta$. See Section 2.3 for details.}
\label{fig:fig1}
\end{figure}

In this paper, we investigate the effects of the interaction parameters on the formation and evolution of CRGs.
In particular, we consider the following parameters (represented schematically in Fig. \ref{fig:fig1}).
\begin{enumerate}
\item The \emph{initial relative velocity} $\mathbf{V}_0$ is the relative velocity of the centre of mass (CM) of the intruder galaxy with respect to the CM of the target galaxy, at the initial distance between the two galaxies.
\item The \emph{impact parameter} $b$ is the component perpendicular to $\mathbf{V}_0$ of the initial distance vector between the CMs of the two galaxies.
\item The \emph{inclination angle} $\vartheta$ is the angle between the symmetry axis of the target disc and $\mathbf{V}_0$.
\end{enumerate}

We run a set of $N$-body/AMR simulations exploring two different values of both $\vartheta$ and $\mathbf{V}_0$, and three different values of $b$.
The initial parameters of our simulation suite are summarized in Table \ref{table2}.
We initially place the CM of the target galaxy at rest in the centre of the box; the disc lies in the $xy$ plane.
The CM of the intruder galaxy has an initial distance $D_{0}$ from the target CM equal to the target halo virial radius, $D_{0} = R_{200}^{\rm (prim)} = 220 \; {\rm kpc}$, and an initial velocity $\mathbf{V}_0$; $\mathbf{V}_0$ lies in the $yz$ plane.
The time elapsed from the beginning of the simulation to the galaxy collision is $300 - 600 \; {\rm Myr}$, depending on $\mathbf{V}_0$; we let the models evolve till $\sim 200 \; {\rm Myr}$ after the interaction time, i.e. the moment of closest approach.
\begin{table}
\caption{List of performed simulations and of their interaction parameters.}
\begin{center}
\label{table2}
\begin{tabular}{l c c c}
\hline
Label & $\vartheta$ & $b$ & $|\mathbf{V}_0|$ \\
 & $[ ^{\circ} ]$ & ${\rm [ kpc ]}$ & ${\rm [ km \, s^{-1} ]}$ \\
\hline
a0b0v25 & 0.0 & 0.0 & 250.0  \\
a0b0v65 & 0.0 & 0.0 & 650.0  \\
a0b5v25 & 0.0 & 5.0 & 250.0  \\
a0b5v65 & 0.0 & 5.0 & 650.0  \\
a0b10v25 & 0.0 & 10.0 & 250.0  \\
a0b10v65 & 0.0 & 10.0 & 650.0  \\
a30b0v25 & 30.0 & 0.0 & 250.0  \\
a30b0v65 & 30.0 & 0.0 & 650.0  \\
a30b5v25 & 30.0 & 5.0 & 250.0  \\
a30b5v65 & 30.0 & 5.0 & 650.0  \\
a30b10v25 & 30.0 & 10.0 & 250.0  \\
a30b10v65 & 30.0 & 10.0 & 650.0  \\
\hline
\end{tabular}
\end{center}
\end{table}

The set up of the encounters ranges from axisymmetric ($b=0 \; {\rm kpc}$, $\vartheta=0^{\circ}$) to highly off-centre and inclined ($b=10 \; {\rm kpc}$, $\vartheta=30^{\circ}$) interactions.
The smaller value for $|\mathbf{V}_{0}| = 250 \; {\rm km \, s^{-1}}$ corresponds to a marginally bound interaction, consistent with observations \citep{madore+09}.
The higher value $|\mathbf{V}_{0}| = 650 \; {\rm km \, s^{-1}}$ represents a very fast interaction within a massive galaxy groups (\citealt{carlberg+01}, \citealt{faltenbacher+07}).


\section{Results}\label{section3}

\subsection{Morphology}

\begin{figure*}
\includegraphics[width=180mm]{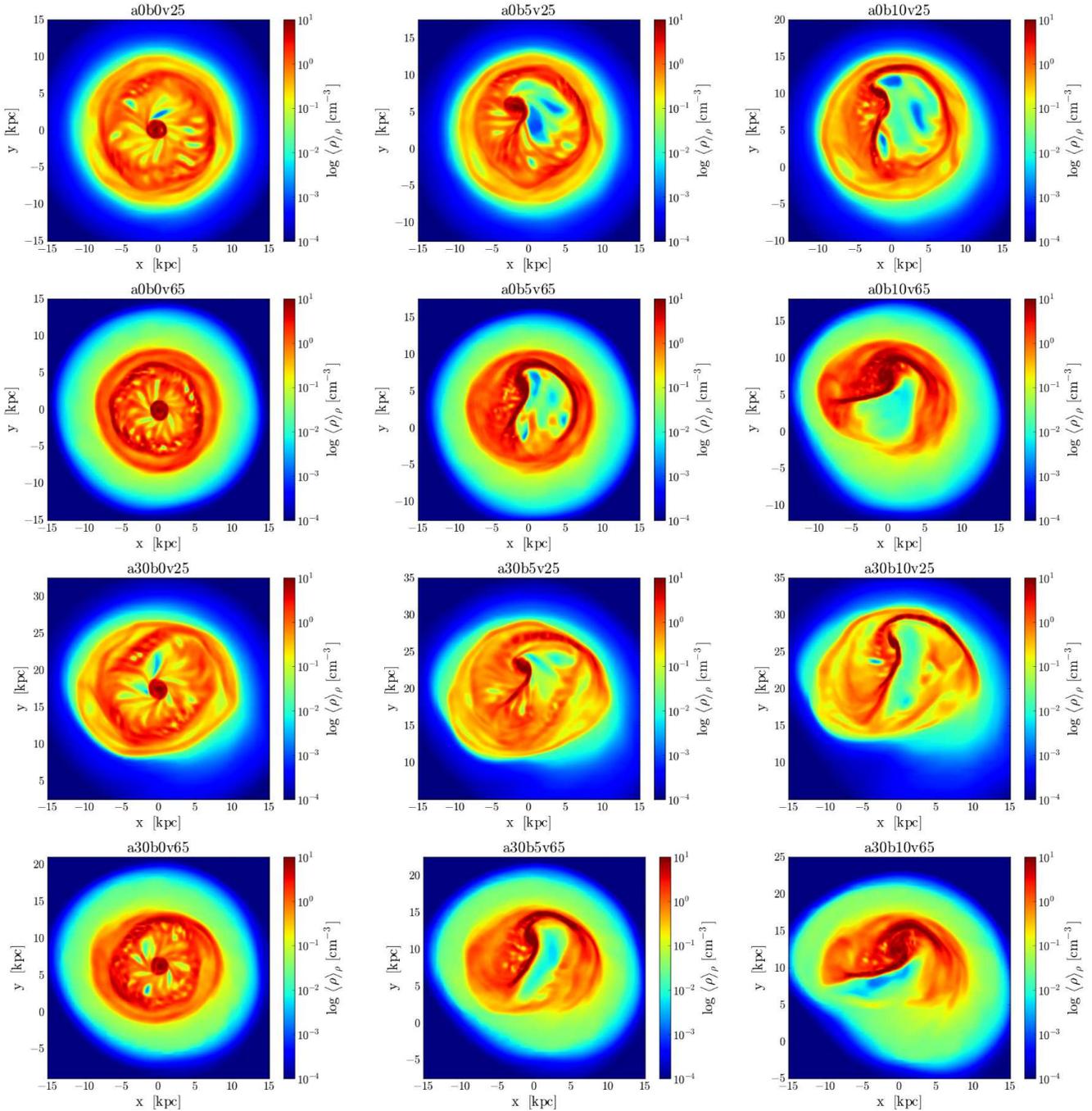}
\caption{Mass-weighted gas density maps in the $xy$ plane for all the runs at time $\tau_{\rm ring} \simeq 50 \; {\rm Myr}$ after the interaction.
From the leftmost to the rightmost column: $b=0 \; {\rm kpc}$, $b=5 \; {\rm kpc}$, $b=10 \; {\rm kpc}$.
The top two rows show interactions with $\vartheta=0^\circ$, while the bottom two rows with $\vartheta = 30^{\circ}$.
For each pair of rows, the first one represents encounters with $|\mathbf{V}_0| = 250 \; {\rm km \, s^{-1}}$, the second one with  $|\mathbf{V}_0| = 650 \; {\rm km \, s^{-1}}$.}
\label{figure_xy}
\end{figure*}

Fig. \ref{figure_xy} shows the face-on view of the simulated target galaxies at time $\tau_{\rm ring} \simeq 50 \; {\rm Myr}$ after the interaction.
Images show the mass-weighted gas density maps.
Regardless of the interaction parameters, the ring-like features of the runs have a typical radius $R_{\rm ring} \simeq 6-10 \; {\rm kpc}$ ($\sim 2-3 R_{\rm d}$).
This value slightly decreases for fast encounters, but the effect is small because the ring expansion velocity is mainly influenced by the mass of the perturber, in agreement with the results of \citet{gerber+96}.

Simulations with $b=0 \; {\rm kpc}$ show circular or nearly elliptical rings, depending on the value of the inclination angle.
The gaseous component of the ring is quite broad (with a width $\Delta R_{\rm ring} \simeq 1.5-2 \; {\rm kpc}$) and clumpy, due to the combined effects of cooling, SF and SN feedback.
Inside the ring we observe radial gaseous spokes that are thicker and slightly less developed for the encounter with $|\mathbf{V}_{0}| = 250 \; {\rm km \, s^{-1}}$, and the presence of a gaseous and star-forming secondary ring that recently started propagating.
Observations of the Cartwheel galaxy do not support our finding that spokes are mainly composed of gas: no HI or H$\alpha$ emission is found in the spokes and in the inner ring (\citealt{higdon+95}, \citealt{higdon+96}).
However, recent observations of NGC 922 \citep{pellerin+10} show H$\alpha$ emission and the presence of young star clusters along the incomplete ring as well as in the central bar and in correspondence to possible spokes.
Also the inner ring of Arp 10 has H$\alpha$ emission and seems to be a site of SF (\citealt*{charmandaris+93}, \citealt{bizyaev+07}).
Therefore, the amount of gas in the spokes and in the central regions of CRGs may depend on the pre-encounter distribution of gas \citep{smith+12}.

Off-centre collisions with $b=5 \; {\rm kpc}$ and $b=10 \; {\rm kpc}$ produce results similar to those reported by \cite{mapelli+12}.
The asymmetry of the interaction induces the displacement of the nucleus in the direction of the impact.
The ring is also asymmetric and composed of two main structures originating from the displaced nucleus and connected at the end: a wide arc and a straighter arm.
The wide arc is generally clumpy (e.g. run a30b5v25 or a30b10v25) and resembles the gravitationally unstable tidal tail produced by galaxy interactions in \citet*{wetzstein+07}.
For simulations with $b=10 \; {\rm kpc}$, the ring is loosely connected for $|\mathbf{V}_{0}| = 250 \; {\rm km \, s^{-1}}$, and not connected in the case of $|\mathbf{V}_{0}| = 650 \; {\rm km \, s^{-1}}$.
This trend is enhanced when the inclination angle $\vartheta$ is increased.
Our results suggest an approximate upper limit $b \lesssim 3 R_{\rm d}$ to produce a connected ring from an off-centre high-speed collision.
Another common feature of all the runs with $b>0 \; {\rm kpc}$ is the development of a stellar and gaseous bar-like feature in correspondence to the displaced nucleus.


\subsubsection{Disc warping} \label{subsub:warping}

\begin{figure*}
\includegraphics[width=180mm]{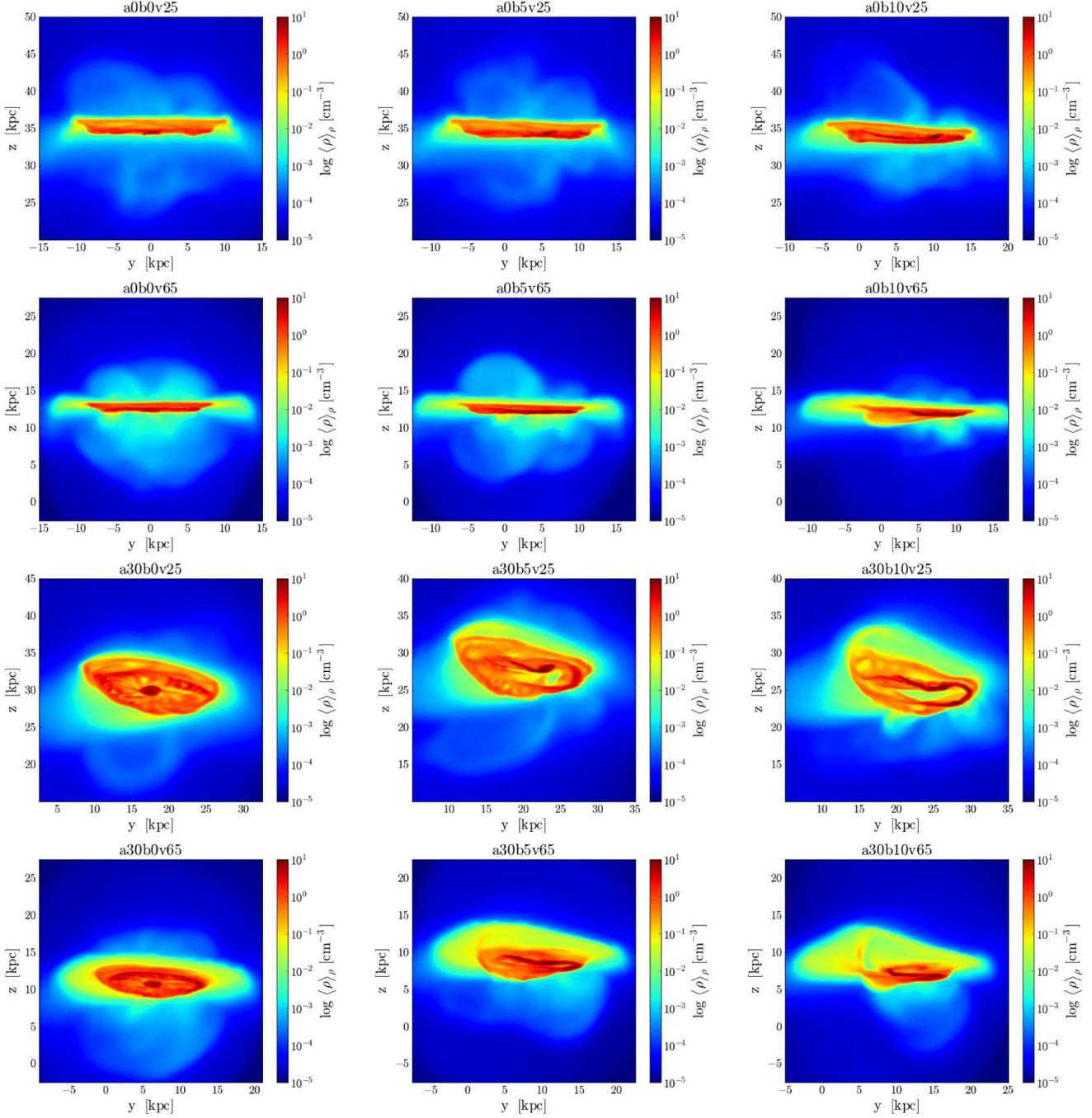}
\caption{Mass-weighted gas density maps in the $yz$ plane of all the runs at time $\tau_{\rm ring} \simeq 50 \; {\rm Myr}$ after the interaction.
From the leftmost to the rightmost column: $b=0 \; {\rm kpc}$, $b=5 \; {\rm kpc}$, $b=10 \; {\rm kpc}$.
The top two rows show interactions with $\vartheta=0^\circ$, while the bottom two rows with $\vartheta = 30^{\circ}$.
For each pair of rows, the first one represents encounters with $|\mathbf{V}_0| = 250 \; {\rm km \, s^{-1}}$, the second one with  $|\mathbf{V}_0| = 650 \; {\rm km \, s^{-1}}$.
}
\label{figure_yz}
\end{figure*}

Fig. \ref{figure_yz} shows edge-on views of all the gaseous target discs at time $\tau_{\rm ring} \simeq 50 \; {\rm Myr}$ after the interaction.
Images show the effects of the interaction on the vertical structure of the target disc.
In all cases, the target is displaced from its original position in the direction of the intruder by $\lesssim 35 \; {\rm kpc}$ and $\lesssim 15 \; {\rm kpc}$ for slower and faster interactions, respectively.
Another common feature is the development of thin gaseous tails in the direction of the intruder passage, especially  for inclined interactions with $\vartheta = 30^{\circ}$.
These tails can be sometimes observed as HI plumes departing from the target disc \citep{higdon+96}.

Off-centre and low-speed interactions produce more disc vertical heating than faster and symmetric encounters.
Moreover, the inclination angle induces a net torque on the target disc along the $x$ axis.
Therefore, after the interaction, the ring develops on a plane that is tilted with respect to the original plane of the disc.
For the case with $b = 0 \; {\rm kpc}$ and $\vartheta > 0^{\circ}$, we can estimate the warp analytically.
Assuming that the intruder moves on a straight line in the $yz$ plane with an inclination angle $\vartheta$ defined as above, only the $x$ component of the torque $\mathbf{N}$ does not vanish on the entire target disc.
It induces an angular momentum variation that can be estimated as (see Appendix \ref{appA} for details):
\begin{eqnarray}
\Delta \mathbf{J}(R,t) & \simeq & \mathbf{e}_{x} \, \int_{-\infty}^{t} {\rm d}\tau \int_{0}^{R} {\rm dr} \, \times \nonumber \\
& & \times  \frac{3 \pi G \, M_{\rm int} ({\rm V} \tau)^2 \, r^3 \, \Sigma(r) \, \sin(2 \vartheta)}{2(r^2+a^2+({\rm V} \tau)^2)^{5/2}}.
\end{eqnarray}
where ${\rm V}$ is the module of the intruder velocity, $a$ is the radial scale of the intruder (for which we assume a Plummer potential, see Appendix \ref{appA}), $M_{\rm int}$ is the intruder baryonic mass and $\Sigma(r)$ is the exponential surface density of the target disc.
The unperturbed component of the angular momentum can be estimated as:
\begin{equation}
\mathbf{J}_0(R) = \mathbf{e}_{z} \, 2 \, \pi \int_{0}^{R} {\rm d}r \, r^2 \Sigma(r) {\rm V}_{\rm circ}(r),
\end{equation}
where ${\rm V}_{\rm circ}(r)$ is the circular velocity of the unperturbed exponential disc.
Then, we can parametrize the warping as:
\begin{equation} \label{eq:angle_warp}
\varphi_{\rm warp}(R,t) = \arctan \left( \frac{|\Delta \mathbf{J}(R,t)|}{|\mathbf{J}_0(R)|} \right).
\end{equation}
$\varphi_{\rm warp}$ is the average warping of the target disc at each radial position, due to the gravitational torque exerted by the encounter.

\begin{figure*}
\begin{minipage}{0.48\textwidth}
\centering
\includegraphics[angle=270,width=80mm]{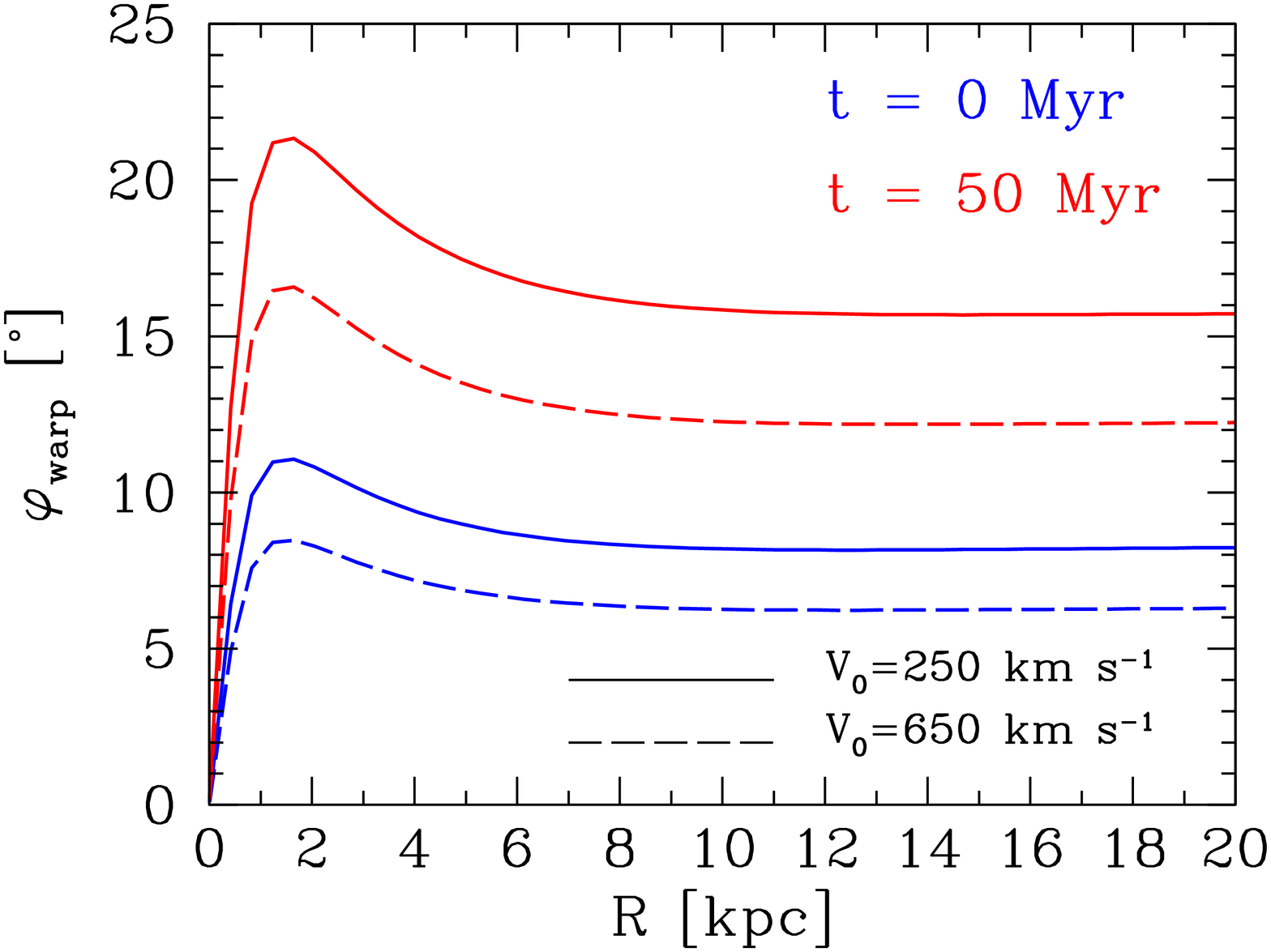}
\end{minipage}
\hfill
\begin{minipage}{0.48\textwidth}
\centering
\includegraphics[width=80mm]{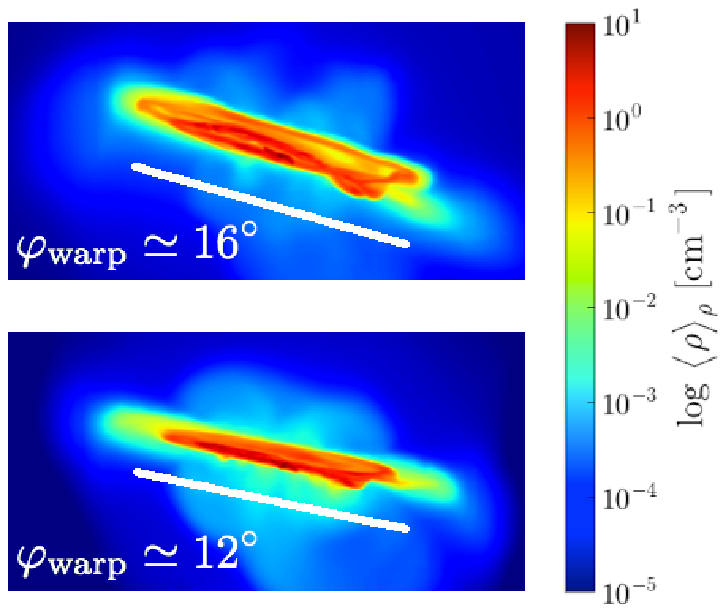}
\end{minipage}
\caption{Left-hand panel: radial profile of $\varphi_{\rm warp}$ as calculated from Eq. (\ref{eq:angle_warp}) at $t = 0 \; {\rm Myr}$ (the time of closest approach), and $t = 50 \; {\rm Myr}$ after the interaction for run a30b0v25 (solid lines) and for run a30b0v65 (dashed lines).
Right-hand panel: gas density maps projected in the $xz$ plane for runs a30b0v25 (top figure) and a30b0v65 (bottom figure).
In both panel, the white line shows the theoretical asymptotic value of $\varphi_{\rm warp}$.
Each map is $40 \times 20 \; {\rm kpc}$ per edge.}
\label{figure_warping}
\end{figure*}

This simple treatment does not take into account the perturbation by the primary disc on the intruder motion and the propagation of the ring\footnote{The impulse approximation adopted to derive equation~(\ref{eq:angle_warp}) is acceptable, as the total mass of the intruder far exceeds the fluctuations induced in the stellar density of the target.}.
Thus, it can only predict the overall inclination of the disc, represented by the nearly constant asymptotic value that $\varphi_{\rm warp}$ assumes for radii $R \gtrsim 7 \; {\rm kpc}$ ($\sim 2 R_{\rm d}$), as shown in the left-hand panel of Fig. \ref{figure_warping}.
This plot shows the radial profile of $\varphi_{\rm warp}$ deduced from Eq. (\ref{eq:angle_warp}) at two characteristic times (0 and 50 Myr after the interaction) for runs a30b0v25 and a30b0v65.
For ${\rm V}$ we use the relative velocity between the CMs of the baryonic components of the two galaxies at the moment of the interaction, as measured from simulations (${\rm V} = 850 \; {\rm km \, s^{-1}}$ for run a30b0v25 and ${\rm V} = 1050 \; {\rm km \, s^{-1}}$ for run a30b0v65).
We assume the radial scale length of the intruder is $a = 0.75 \; {\rm kpc}$ (see Appendix \ref{appA}).
For run a30b0v25, $\varphi_{\rm warp}$ approaches the asymptotic value $\varphi_{\rm warp} \simeq 16^{\circ}$, while for run a30b0v65 $\varphi_{\rm warp} \simeq 12^{\circ}$.
These values are in fair agreement with simulations (as shown in the right-hand panel of Fig. \ref{figure_warping}) and represent the overall inclination of the target disc with respect to the $xy$ plane.

Off-centre interactions with $\vartheta > 0^{\circ}$ produce rings completely warped in the vertical direction, especially for slow encounters (see Fig. \ref{figure_yz}).
In these cases, the loosely connected ring develops with a nearly helicoidal shape because of the coupled role played by the inclination angle, which induce the vertical torque discussed above, and by the asymmetric propagation of the expanding ring, due to $b > 0 \; {\rm kpc}$ \citep{struck-marcell+90}.


\subsection{SF history}

\begin{figure*}
\includegraphics[angle=270,width=170mm]{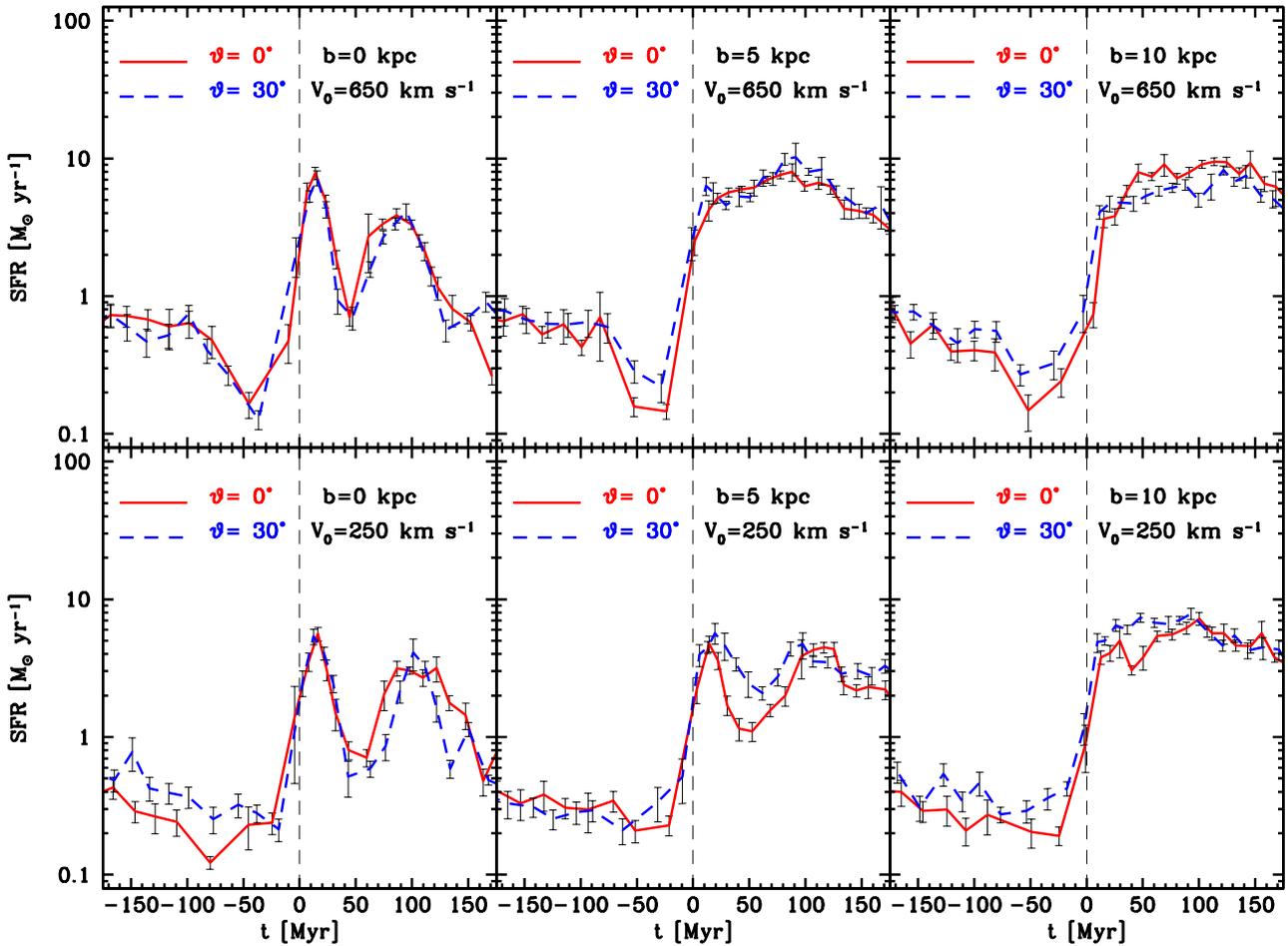}
\caption{Effect of the inclination angle $\vartheta$ on the SF history.
Each panel shows the SFR as a function of time, where $t=0 \; {\rm Myr}$ refers to the interaction time; the solid (dashed) line refers to the case with $\vartheta = 0^{\circ}$ ($\vartheta = 30^{\circ}$).
The top row shows runs with $|\mathbf{V}_0| = 650 \; {\rm km \, s^{-1}}$, while the bottom row shows runs with $|\mathbf{V}_0| = 250 \; {\rm km \, s^{-1}}$.
From the leftmost to the rightmost column: runs with $b=0 \; {\rm kpc}$, $b=5 \; {\rm kpc}$ and $b=10 \; {\rm kpc}$, respectively.
Error bars represent 1-$\sigma$ uncertainty on the estimated SFR.
}
\label{figure_SFRa}
\end{figure*}

\begin{figure*}
\includegraphics[angle=270,width=170mm]{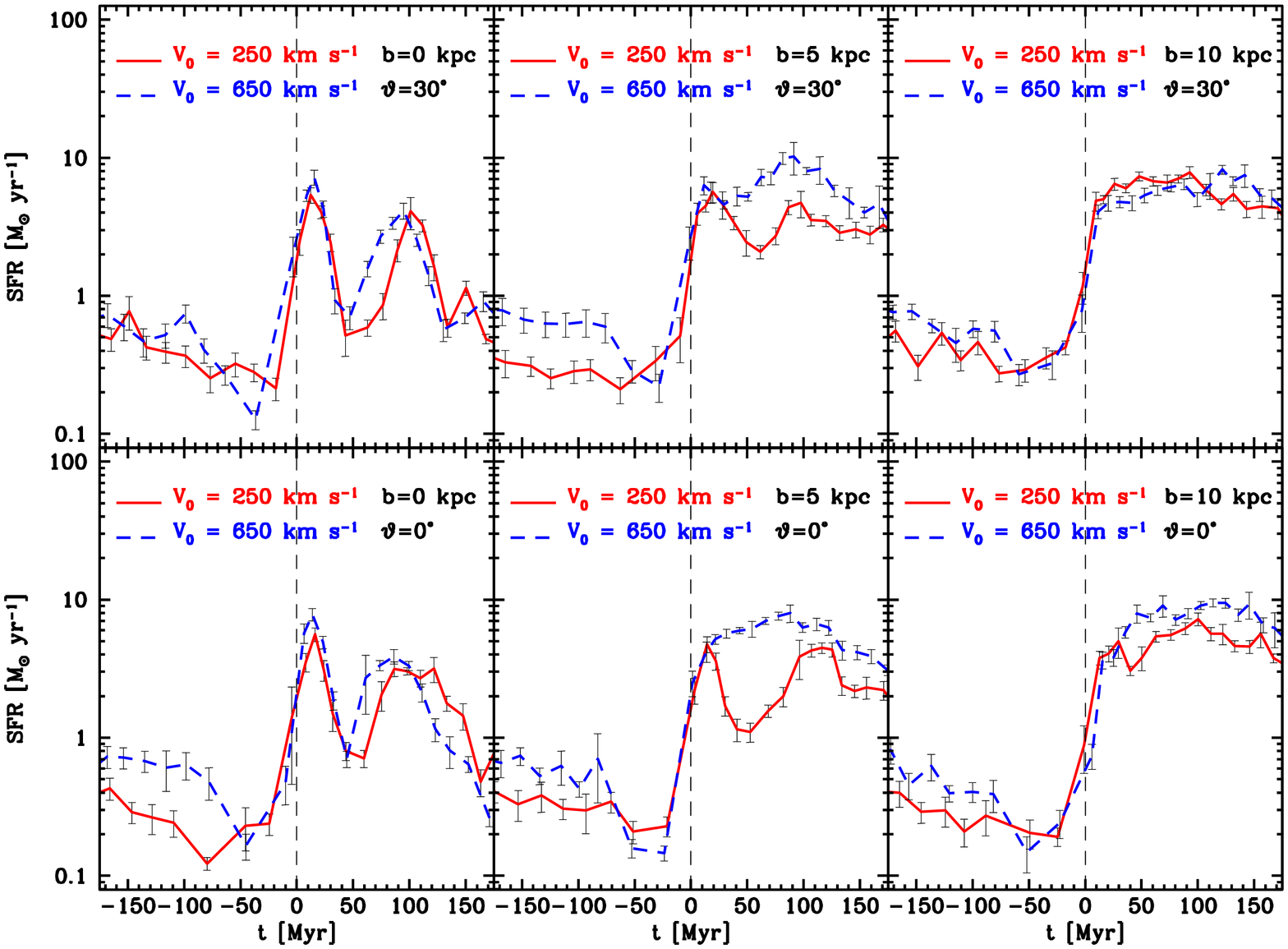}
\caption{Effect of the initial relative velocity $|\mathbf{V}_0|$ on the SF history.
Each panel shows the SFR as a function of time, where $t=0 \; {\rm Myr}$ refers to the interaction time; the solid (dashed) line refers to the case with $|\mathbf{V}_0| = 250 \; {\rm km \, s^{-1}}$ ($|\mathbf{V}_0| = 650 \; {\rm km \, s^{-1}}$).
The top row shows runs with $\vartheta = 30^{\circ}$, while the bottom row shows runs with $\vartheta = 0^{\circ}$.
From the leftmost to the rightmost column: runs with $b=0 \; {\rm kpc}$, $b=5 \; {\rm kpc}$, $b=10 \; {\rm kpc}$, respectively.
Error bars represent 1-$\sigma$ uncertainty on the estimated SFR.
}
\label{figure_SFRv}
\end{figure*}

We study the effects of the interaction parameters on the global SF histories of our simulations.
In Fig. \ref{figure_SFRa}, we show the SFR as a function of time for all the runs with different inclination angle $\vartheta$, while in Fig. \ref{figure_SFRv} we compare the SF histories for runs with different initial relative velocity $\mathbf{V}_0$.
Before interaction, the target has a low ${\rm SFR} \sim 0.1-1 \; {\rm M_{\odot} \, yr^{-1}}$, consistent with the observational Kennicutt-Schmidt relation \citep{kennicutt+98}.
For all models, the target experiences an enhancement of SF when the intruder crosses the target disc.
The SFR rapidly increases from $\lesssim 1 \; {\rm M_{\odot} \, yr^{-1}}$ in the pre-encounter phase to $\sim 10 \; {\rm M_{\odot} \, yr^{-1}}$ during the starburst, in agreement with observations (\citealt{mayya+05}, \citealt{rappaport+10}, \citealt{fogarty+11}).
Figs. \ref{figure_SFRa} and \ref{figure_SFRv} suggest that neither $\vartheta$ nor $\mathbf{V}_{0}$ influence significantly the SF history of a typical CRG.

\begin{figure*}
\includegraphics[angle=270,width=170mm]{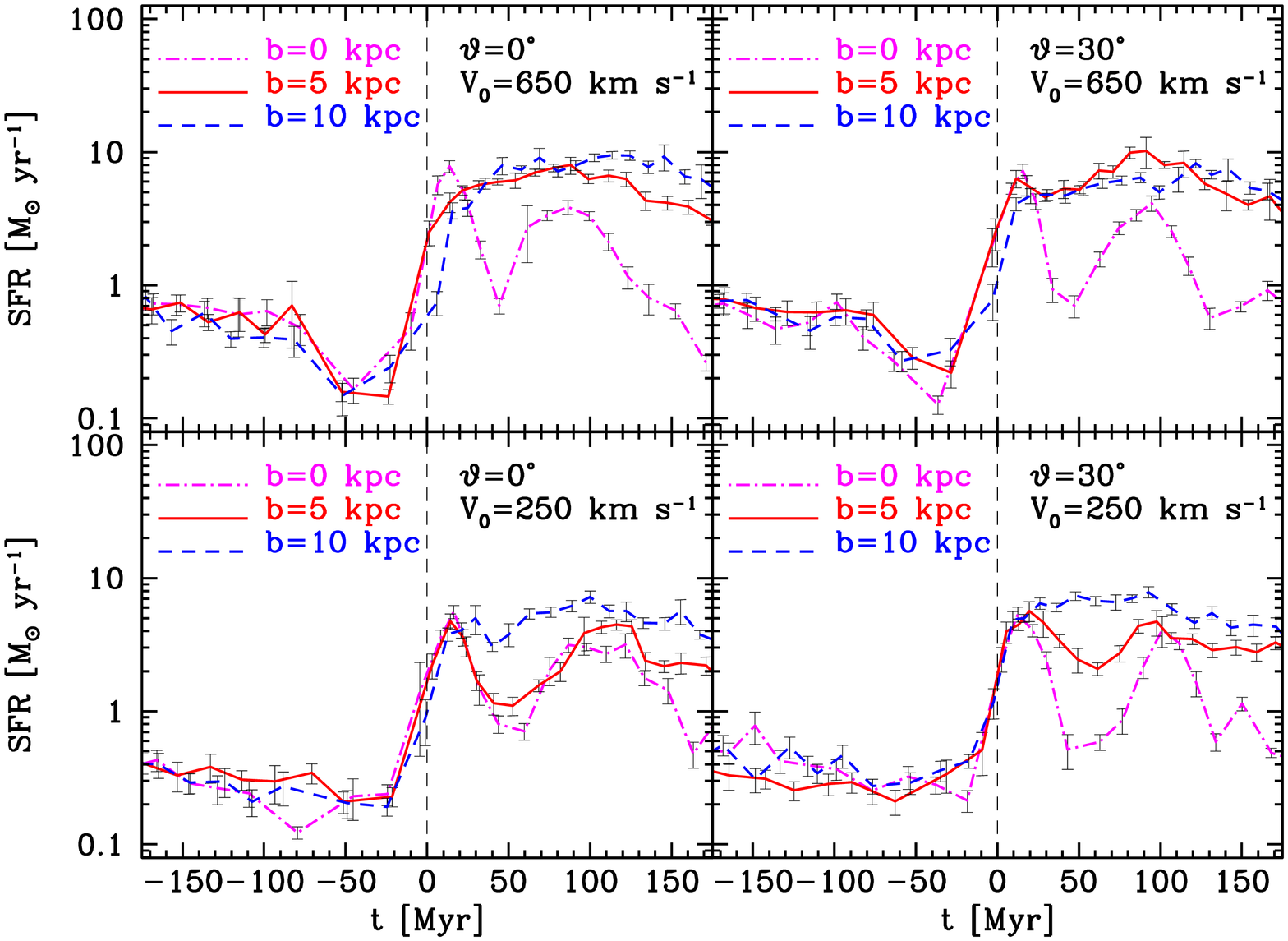}
\caption{Effect of the impact parameter $b$ on the SF history.
Each panel shows the SFR as a function of time, where $t=0 \; {\rm Myr}$ refers to the interaction time; the dot-dashed, solid and dashed lines refer to the cases with $b = 0 \; {\rm kpc}$, $b = 5 \; {\rm kpc}$ and $b = 10 \; {\rm kpc}$, respectively.
The top row shows runs with $|\mathbf{V}_0| = 650 \; {\rm km \, s^{-1}}$, while the bottom row shows runs with $|\mathbf{V}_0| = 250 \; {\rm km \, s^{-1}}$.
The left and right columns show runs with $\vartheta=0^{\circ}$ and $\vartheta=30^{\circ}$, respectively.
Error bars represent 1-$\sigma$ uncertainty on the estimated SFR.
}
\label{figure_SFRb}
\end{figure*}

By contrast, the impact parameter $b$ influences significantly the SF history in our simulations, as shown in Fig. \ref{figure_SFRb}.
We systematically observe a double-peaked SF history for axisymmetric encounters.
After the interaction, the SFR steeply increases, reaches a maximum and then decreases on a typical time scale $\tau_{\rm SF} \sim 50 \; {\rm Myr}$.
This is probably because the ring travels rapidly across the disc during the first $\sim 100 \; {\rm Myr}$ after the interaction; when it reaches the outskirts of the disc, the lower gas density quenches the SF.
The first SF bump is followed by a second episode of SF due to the propagation of the secondary ring.

Instead, after the initial starburst, all off-centre encounters show a nearly constant or slowly rising SFR that lasts more than $\sim 100 \; {\rm Myr}$.
The SF histories of runs with $b = 5 \; {\rm kpc}$ and $|\mathbf{V}_0| = 250 \; {\rm km \, s^{-1}}$ represent a transition between truly axisymmetric and strongly off-centre interactions, as they still show a less pronounced double-peaked shape.

These results are in agreement with \citet{mapelli+12}.
In particular, the SFR in their C and D models (with $f_{\rm gas} \simeq 0.09$ and $f_{\rm gas} \simeq 0.05$, respectively) reaches $\sim 20 \; {\rm M_{\odot} \, yr^{-1}}$ and $\sim 8 \; {\rm M_{\odot} \, yr^{-1}}$, respectively, during the first $50 \; {\rm Myr}$ after the interaction.
Our run a0b10v65 is qualitatively comparable to theirs and we find similar values of ${\rm SFR} \sim 8-10 \; {\rm M_{\odot} \, yr^{-1}}$ despite the different implementation of the SF in the adopted codes.

None of the considered interaction parameters appreciably affects the peak of SFR.
This feature is more sensitive to other parameters such as $f_{\rm gas}$, as pointed out by \citet{mapelli+12}.


\subsection{Synthetic photometry}

Photometric properties can be inferred from our simulations, by using synthetic stellar population models.
We use mass-to-light ratio tables\footnote{See {\tt http://stev.oapd.inaf.it/cgi-bin/cmd}.} based on the isochrones and synthetic stellar populations of \citet{marigo+08} and \citet{girardi+10}.
These tables span the stellar age interval from $4 \times 10^6 \; {\rm yr}$ to $12.6 \times 10^{10} \; {\rm yr}$ and give the magnitude in the $UBVRIJHK$ bands for a stellar system with total mass $1 \; {\rm M_{\odot}}$ and a Chabrier initial mass function (IMF; \citealt{chabrier+01}).
We assume a constant sub-solar metallicity $Z = 0.0019 = 0.1 Z_{\odot}$, in agreement with observations of CRGs (\citealt{fosbury+77}; \citealt{bransford+98}).

\begin{figure*}
\includegraphics[width=180mm]{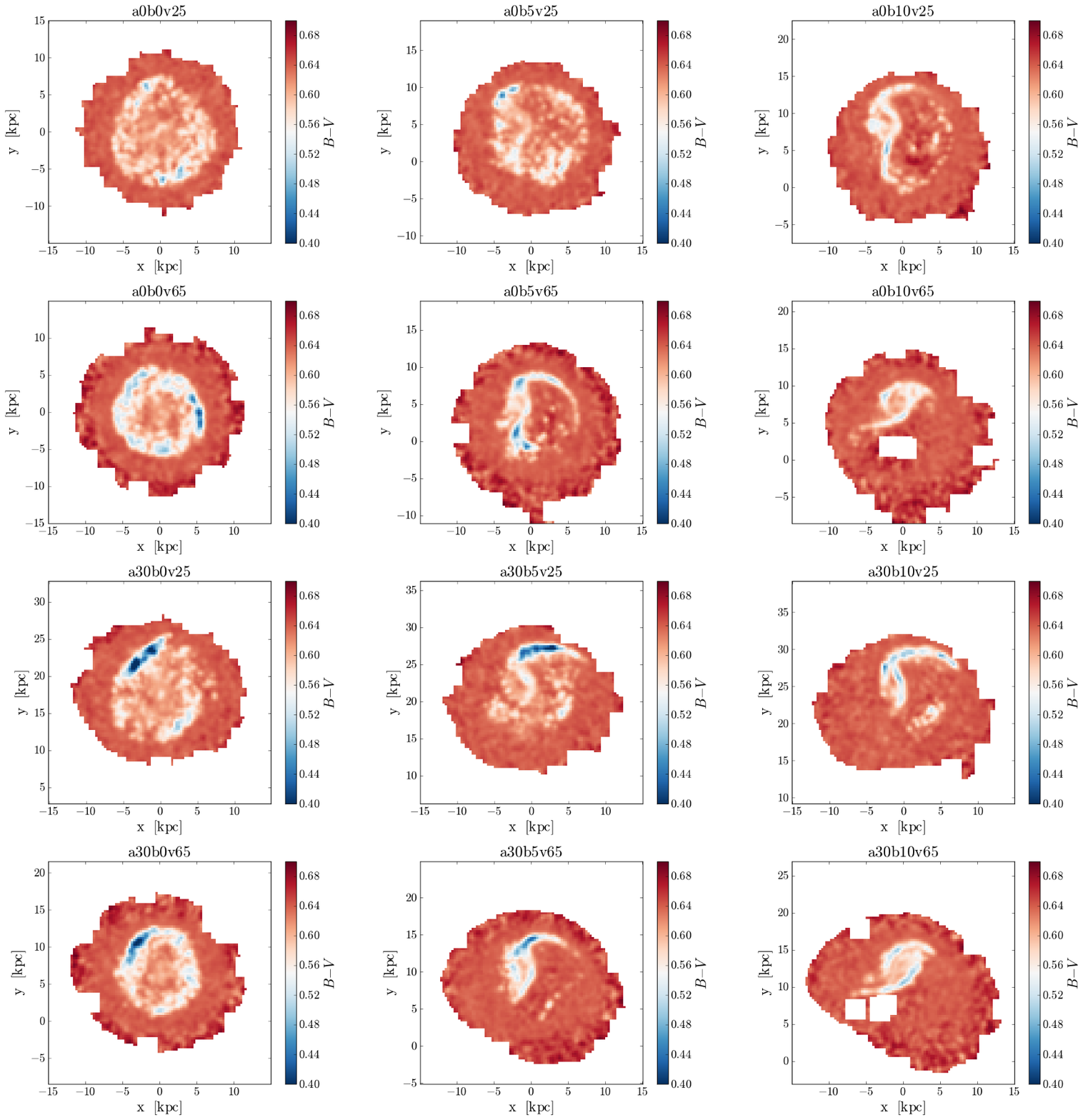}
\caption{
$B-V$ colour maps of all the simulations in the $yz$ plane at time $\tau_{\rm ring} \simeq 50 \; {\rm Myr}$ after the interaction.
Maps are computed assuming a Chabrier IMF and a constant metallicity $Z=0.1 Z_{\odot}$.
From the leftmost to the rightmost column: $b=0 \; {\rm kpc}$, $b=5 \; {\rm kpc}$ and $b=10 \; {\rm kpc}$.
The top two rows show interactions with $\vartheta=0^\circ$, while the bottom two rows with $\vartheta = 30^{\circ}$.
For each pair of rows, the first one represents encounters with $|\mathbf{V}_0| = 250 \; {\rm km \, s^{-1}}$, the second one with  $|\mathbf{V}_0| = 650 \; {\rm km \, s^{-1}}$.
}
\label{figure_B-V}
\end{figure*}

In Fig. \ref{figure_B-V} the synthetic $B-V$ colour maps obtained for all our simulations at time $\tau_{\rm ring} \simeq 50 \; {\rm Myr}$ are shown.
We calculated these maps after assigning a random age (between $1 \; {\rm Gyr}$ and $10 \; {\rm Gyr}$) to stellar particles that were in the ICs; in this process we do not take into account any pre-interaction age/metallicity gradient.
For purpose of visualization, maps are smoothed with a Gaussian filter with $\sigma = 300 \; {\rm pc}$.

All models show a blue ring-like feature with typical colour $(B-V)_{\rm ring} \sim 0.5 \; {\rm mag}$.
The rings are sprinkled with bluer knots corresponding to regions of recent SF, with colour $(B-V)_{\rm knot} \lesssim 0.4 \; {\rm mag}$.
The underlying disc is redder with an average colour $(B-V)_{\rm disc} \sim 0.65 \; {\rm mag}$.
The ring is characterized by a blue colour bump with respect to the disc, $\Delta (B-V) \sim 0.2 \; {\rm mag}$, which is consistent with $B-V$ profiles derived by \citet*{romano+08} for a sample of 15 CRGs.
We repeated the same analysis assuming a metallicity $Z=0.019=Z_{\odot}$, and we found similar results, even if with a redder ($\sim 0.15-0.2 \; {\rm mag}$) overall colour.

Simulations with $b = 0 \; {\rm kpc}$ show a gaseous secondary ring that just started to expand (see Fig. \ref{figure_xy}) and that is expected to form stars, but colour maps do not show an excess of blue colour in the inner regions of the ring.
This is in partial agreement with observations: for example, the Cartwheel galaxy shows a red but gas-poor secondary ring (\citealt{higdon+95}, \citealt{higdon+96}).
In our models, the secondary ring starts forming stars $\sim 50 \; {\rm Myr}$ after, in correspondence to the second peak of the SFR at $\sim 100 \; {\rm Myr}$ after the interaction.

Off-centre collisions show very blue knots mainly located in the wide arc and at the extremity of the bar-like nuclear region.
This suggests that the asymmetry of the encounter can funnel gas to the nucleus and favor SF near it, as observed, for example, in NGC 922 \citep{pellerin+10}.

Our simulations likely underestimate the clumpiness of the blue knots in the ring, as we cannot resolve the giant molecular clouds and thus we cannot produce realistic SF agglomerates, such as young dense star clusters (see Section \ref{subsec21} for a discussion about the limitations of the model).

\section{Summary}\label{section4}

In this paper, we investigate the properties of CRGs as a function of the interaction parameters by means of AMR hydrodynamical simulations.
We focus on three parameters: the impact parameter, the initial relative velocity and the inclination angle.

All the simulations show a few common features, such as the radius of the ring ($R_{\rm ring} \simeq 6-10 \; {\rm kpc}$) at time $\tau_{\rm ring} \simeq 50 \; {\rm Myr}$ after the interaction, which is weakly influenced by the three interaction parameters described above.
On the other hand, axisymmetric encounters produce CRGs noticeably different with respect to those formed by off-centre interactions.
Axisymmetric interactions result in a circular primary ring followed by a smaller secondary ring, whereas CRGs born from off-centre collisions show asymmetric rings and a displaced nucleus.
In the most asymmetric cases ($b = 10 \; {\rm kpc}$ and $\vartheta = 30^{\circ}$) the ring is not connected.
The maximum impact parameter to form a  connected ring is $b \lesssim 3 R_{\rm d}$, where $R_{\rm d}$ is the scale length of the target disc.

Interactions with $\vartheta > 0^{\circ}$ induce a torque on the target disc producing a vertical warp.
In the special case of encounters with $b=0 \; {\rm kpc}$ and $\vartheta > 0^{\circ}$, the ring develops on a plane tilted with respect to the original target disc plane and we propose a simple analytical treatment to predict the overall warping of the target disc after the interaction.

After the interaction, the ring expansion enhances the SF in the target disc, reaching a typical SFR $\sim 10 \; {\rm M_{\odot} \, yr^{-1}}$, in agreement with observations (\citealt{higdon+97}, \citealt{fogarty+11}).
We show that the initial velocity and the inclination angle have weak effects on the peak of the starburst and on the evolution with time of the global SFR.
Instead, the impact parameter has a crucial effect on SF: symmetric encounters cause a short-lived starburst episode, whereas off-centre interactions produce long-lasting ($\gtrsim 100 \; {\rm Myr}$) SF events.

The expanding ring is the region mainly affected by SF, as suggested by observations (e.g. \citealt{higdon+97}, \citealt{romano+08}).
This is confirmed by the synthetic colour maps (based on the models by \citealt{marigo+08} and \citealt{girardi+10}) that we compute from the simulations.
We find an average $B-V$ colour $\sim 0.5 \; {\rm mag}$ for the ring, with clumps of massive SF with $B-V \lesssim 0.4 \; {\rm mag}$.
The ring is typically $\sim 0.1-0.2 \; {\rm mag}$ bluer than the rest of the galaxy, in agreement with the $B-V$ colour profiles measured by \citet{romano+08} for a sample of 15 CRGs.

A number of issues cannot be explained by our simulations.
For example, it is still unclear which mechanism regulates SF in the inner regions of the ring, since HI/H$\alpha$ observations suggest that secondary rings are gas-poor (e.g., the Cartwheel, \citealt{higdon+95}; \citealt{higdon+96}; but see also \citealt{bizyaev+07} about Arp 10), while our simulations predict the opposite.
This may be due to several reasons, such as the initial distribution of gas inside the disc, or the idealized sub-grid SF process that does not take into account the chemical abundances and the role of $\rm H_2$.
Thus, future works will require more realistic initial gas distributions and/or more detailed recipes of SF (e.g. $\rm H_2$ regulated SF, \citealt*{krumholz+09}; \citealt{kuhlen+12}) to better constrain the SF in CRGs.
Another interesting issue we did not explore (but plan to investigate in the future) is the effect of a gas-rich intruder.
In this respect, we remind that the SPH simulations of \citet{struck+97} suggest that the main differences between encounters with (disc-like) gas-rich and gas-poor intruders are related to the SF history, whereas the effects on the ring morphology are quite limited, even if some features (such as plumes) appear to be more developed in the gas-rich case.


\section*{Acknowledgments}

We thank the anonymous referee for comments that significantly improved the paper.
We thank R. Teyssier for useful advice about the use of RAMSES.
We thank L. Widrow for providing us the code to generate the initial conditions.
We thank L. Girardi for maintaining his public website about stellar isochrones.
We thank the authors of PyMSES, which we used for analysing the simulations.
All the simulations were performed with the SP6 cluster at the Consorzio interuniversitario dell'Italia Nord-Est per il Calcolo Automatico (CINECA).
We acknowledge the CINECA Award N. HP10CLI3BX and HP10B3BJEW, 2011 for the availability of high performance computing resources and support.
We acknowledge financial support from INAF through grant PRIN-2011-1.
We thank L. Mayer for useful discussions.



\appendix

\section{Analytical calculation of the warping} \label{appA}

We model the intruder galaxy as a Plummer sphere with potential:
\begin{equation}
\Phi_{\rm int}(r) = - \frac{G M_{\rm int}}{a} \frac{1}{\sqrt{1+(r/a)^{2}}},
\end{equation}
where $a$ is the scale radius of the potential and $M_{\rm int}$ is the intruder baryonic mass\footnote{We neglect the role of the DM halos, for simplicity.}.
We assume that the intruder moves on a straight line in the $yz$ plane, according to $\mathbf{R}(t) = (0, -{\rm V} \, t \, \sin \vartheta, -{\rm V} \, t \, \cos \vartheta)$, where $\vartheta$ is the inclination angle and ${\rm V}$ is intruder velocity.
For simplicity, we assume that ${\rm V}$ is constant in time.
The target disc initially lies on the $xy$ plane and has a surface density profile $\Sigma(r) = \frac{M{\rm d}}{2 \pi R_{\rm d}^2} e^{-r/R_{\rm d}}$.
The force exerted on an infinitesimal mass portion ${\rm d}m = \Sigma(r)\, r \, {\rm d}r \, {\rm d}\phi$ located at $\mathbf{x} = (r \cos \phi, r \sin \phi , 0)$  of the target disc is:
\begin{eqnarray}
{\rm d}\mathbf{F}(r,\phi,t) & = & -{\rm d}m \nabla \Phi_{\rm int} (|\mathbf{x}-\mathbf{R}(t)|)= \nonumber \\
& = &  - \frac{\Sigma(r)\, r \, {\rm d}r \, {\rm d}\phi \, G M_{\rm int}}{(r^2+({\rm V} t)^2 + 2 {\rm V} \, t \, r \sin \vartheta \sin \phi + a^2  )^{3/2}} \times \nonumber \\
& & \times 
\left(
 \begin{array}{c}
r \cos \phi \\
r \sin \phi + {\rm V} \, t \, \sin \vartheta \\
{\rm V} \, t \, \cos \vartheta
\end{array}
\right).
\end{eqnarray} 
We can now derive the torque exerted on the same mass element:
\begin{eqnarray} \label{eq:dN_A}
{\rm d}\mathbf{N}(r,\phi,t) & = & \mathbf{x} \wedge {\rm d}\mathbf{F}(r,\phi,t) = \nonumber \\
 & = & -\frac{G M_{\rm int} \, r \, {\rm V} \, t}{(r^2 + a^2 +2 ({\rm V} t)^2)^{3/2}} \frac{\Sigma(r)\, r \, {\rm d}r \, {\rm d}\phi}{(1+C(r,t) \sin \phi)^{3/2}} \times \nonumber \\
& & \times \left( 
\begin{array}{c}
\cos \vartheta \, \sin \phi \\
-\cos \vartheta \, \cos \phi \\
\sin \vartheta \, \cos \phi \\
\end{array}
\right),
\end{eqnarray}
where:
\begin{equation}
C(r,t) = \frac{2 {\rm V} \, t \, r \sin \vartheta}{r^2+a^2+({\rm V} t)^2}.
\end{equation}
Integrating Eq. (\ref{eq:dN_A}) over ${\rm d}\phi$, we obtain the total torque exerted on an infinitesimally thin annulus of radius $r$.
We point out that the $y$ and $z$ components are proportional to:
\begin{equation}
\int_{0}^{2\pi}  \frac{\cos \phi \, {\rm d}\phi}{(1+C(r,t) \sin \phi)^{3/2}} = 0.
\end{equation}
Therefore, the resulting torque per unit length is:
\begin{eqnarray}
\frac{{\rm d} \mathbf{N}(r,t)}{{\rm d}r} & = & -\frac{G M_{\rm int} \, r^2 \, {\rm V} \, t \, \Sigma(r) \cos \vartheta}{(r^2 + a^2 +2 ({\rm V} t)^2)^{3/2}} \times \nonumber \\
& & \times \, \mathbf{e}_{x} \int_{0}^{2 \pi} \frac{\sin \phi \, {\rm d} \phi}{(1+C(r,t) \sin \phi)^{3/2}}.
\end{eqnarray}
Integrating this equation with the boundary condition $\mathbf{N}(R=0) = 0$, we obtain the total torque:
\begin{eqnarray} \label{eq:N_A}
\mathbf{N}(R,t) & = & - \int_{0}^{R} {\rm d}r \frac{G M_{\rm int} \, r^2 \, {\rm V} \, t \, \Sigma(r) \cos \vartheta}{(r^2 + a^2 +2 ({\rm V} t)^2)^{3/2}} \times \nonumber \\
& & \times \, \mathbf{e}_{x} \int_{0}^{2 \pi} \frac{\sin \phi \, {\rm d} \phi}{(1+C(r,t) \sin \phi)^{3/2}}.
\end{eqnarray}
The angular momentum $\mathbf{J}$ is related to the torque $\mathbf{N}$ by:
\begin{equation}
\frac{{\rm d} \mathbf{J}(R,t)}{{\rm d} t} = \mathbf{N}(R,t).
\end{equation}
We can integrate in time Eq. (\ref{eq:N_A}) to obtain the variation of $\mathbf{J}$ due to the torque exerted by the intruder, $\Delta \mathbf{J}$, with the condition $\lim_{t \rightarrow -\infty} \Delta \mathbf{J} = 0$ (i.e. we assume that for $t \rightarrow -\infty$ $|\mathbf{R}| = +\infty$ and the torque is zero).
We finally obtain:
\begin{eqnarray}
\Delta \mathbf{J}(R,t)  & = & \int_{-\infty}^{t} {\rm d}\tau \, \mathbf{N}(R,t) = \nonumber \\
& = & - \int_{-\infty}^{t} {\rm d}\tau \int_{0}^{R} {\rm dr} \frac{G \, M_{\rm int} {\rm V} \, \tau \, r^2 \, \Sigma(r) \, \cos \vartheta }{(r^2+a^2+({\rm V} \tau)^2)^{3/2}} \times \nonumber \\
& & \times \, \bmath{e}_{x} \int_{0}^{2 \pi} {\rm d} \phi \frac{\sin \phi}{(1+C(r,\tau) \sin \phi)^{3/2}}.
\end{eqnarray}
This expression can be simplified by assuming $\sin \vartheta \ll 1$.
In this case we have:
\begin{eqnarray} \label{eq:Dj_approx}
\Delta \mathbf{J}(R,t) & \simeq & \mathbf{e}_{x} \, \int_{-\infty}^{t} {\rm d}\tau \int_{0}^{R} {\rm dr} \, \times \nonumber \\
& & \times  \frac{3 \pi G \, M_{\rm int} ({\rm V} \tau)^2 \, r^3 \, \Sigma(r) \, \sin(2 \vartheta)}{2(r^2+a^2+({\rm V} \tau)^2)^{5/2}}.
\end{eqnarray}

In this treatment, the $z$ component of the angular momentum $\mathbf{J}$ is not perturbed by the passage of the intruder.
The unperturbed component of the angular momentum is given by:
\begin{equation}
\mathbf{J}_0(R) = \mathbf{e}_{z} \, 2 \, \pi \int_{0}^{R} {\rm d}r \, r^2 \Sigma(r) {\rm V}_{\rm circ}(r),
\end{equation}
where the circular velocity ${\rm V}_{\rm circ}$ for an exponential disc is given by \citep{binney+08}:
\begin{equation}
{\rm V}_{\rm circ}(R) = \sqrt{\frac{2 G M_{\rm d}}{R_{\rm d}}} y \left[ I_0(y) K_0(y) - I_1(y) K_1(y) \right]^{1/2},
\end{equation}
where $y=R/(2 R_{\rm d})$ and $I_0$,$I_1$,$K_0$,$K_1$ are modified Bessel functions.
The warping of the disc can be parametrized by the warping angle $\varphi_{\rm warp}$, defined as:
\begin{equation}
\varphi_{\rm warp}(R,t) = \arctan \left( \frac{|\Delta \mathbf{J}(R,t)|}{|\mathbf{J}_0(R)|} \right).
\end{equation}
This angle represents the inclination of the disc with respect to the $xy$ plane around the $y$ axis at each radius $R$ at time $t$ due to the gravitational torque exerted by the encounter.

This treatment of the torque induced by the intruder galaxy on the target disc is very approximated, as it does not take into account the gravitational effects of the target galaxy on the motion of the intruder and the propagation of the ring as a consequence of the interaction.
Moreover, the intruder is modeled as a Plummer sphere for the sake of simplicity.
However, we choose the radial scale $a$ such that the mass enclosed inside $a$ by the Plummer profile is the same as enclosed by the Hernquist profile:
\begin{equation}
a = \frac{a_{\rm s}}{\sqrt{2^{3/2}-1}} \simeq 0.7495 \, a_{\rm s},
\end{equation}
where $a_{\rm s}$ is the scale radius of the Hernquist profile.
Despite the approximations, this treatment roughly highlights the effects of a non-axisymmetric interaction with inclination angle $\vartheta > 0^{\circ}$ (see \S \ref{subsub:warping}).

\label{lastpage}

\end{document}